\documentstyle[twoside,fleqn,espcrc2,epsf]{article}

\newcommand{\be}{\begin{equation}}
\newcommand{\ee}{\end{equation}}
\newcommand{\bdm}{\begin{displaymath}}
\newcommand{\edm}{\end{displaymath}}

\def\spose#1{\hbox to 0pt{#1\hss}}
\def\ltapprox{\mathrel{\spose{\lower 3pt\hbox{$\mathchar"218$}}
 \raise 2.0pt\hbox{$\mathchar"13C$}}}
\def\gtapprox{\mathrel{\spose{\lower 3pt\hbox{$\mathchar"218$}}
 \raise 2.0pt\hbox{$\mathchar"13E$}}}
\def\inapprox{\mathrel{\spose{\lower 3pt\hbox{$\mathchar"218$}}
 \raise 2.0pt\hbox{$\mathchar"232$}}}

\title{
Overlap Dirac Operator, Eigenvalues and Random Matrix Theory
}

\author{
Robert G.~Edwards$^{\rm \ a}$\thanks{Presented by R.G.~Edwards},
Urs M.~Heller
\address{
SCRI, Florida State University, 
Tallahassee, FL 32306-4130, USA},
Joe Kiskis
\address{
Dept. of Physics, University of California,
Davis, CA 95616, USA},
Rajamani Narayanan
\address{
American Physical Society,
One Research Road,
Ridge, NY 11961, USA}
}

\begin{document}

\begin{abstract}
The properties of the spectrum of the overlap Dirac operator and their
relation to random matrix theory are studied.  In particular, the
predictions from chiral random matrix theory in topologically
non-trivial gauge field sectors are tested.
\end{abstract}

\maketitle

An important property of massless QCD is the spontaneous breaking of
chiral symmetry. The associated Goldstone pions dominate the
low-energy, finite-volume scaling behavior of the Dirac operator
spectrum in the microscopic regime, $1/\Lambda_{QCD} << L << 1/m_\pi$,
with $L$ the length of the system~\cite{LS}. This behavior can be
characterized by chiral random matrix theory (RMT). The RMT
description of the low-energy, finite-volume scaling behavior is
specified by symmetry properties of the Dirac operator and the
topological charge sector being considered~\cite{SV,RMTreview}. The
RMT predictions are universal in the sense that the symmetry
properties, but not the form of the potential matters
~\cite{univ}. Furthermore, the properties can be derived directly from
the effective, finite-volume partition functions of QCD of Leutwyler
and Smilga, without the detour through RMT~\cite{RMTreview}, though
RMT nicely and succinctly describes and classifies all these
properties. The topological charge enters the RMT prediction via the
number of fermionic zero modes, related to the topological charge
through the index theorem. The symmetry properties of the Dirac
operator fall into three classes, corresponding to the chiral
orthogonal, unitary, and symplectic ensembles~\cite{RMTreview}.
Examples are, respectively, fermions in the fundamental representation
of gauge group SU(2), fermions in the fundamental representation of
gauge group SU($N_c$) with $N_c \ge 3$, and fermions in the adjoint
representation of gauge group SU($N_c$).

\begin{figure*}[t]
\vspace*{-10mm} \hspace*{-0cm}
\begin{center}
\epsfxsize = 0.8\textwidth
\centerline{\epsfbox[100 215 550 500]{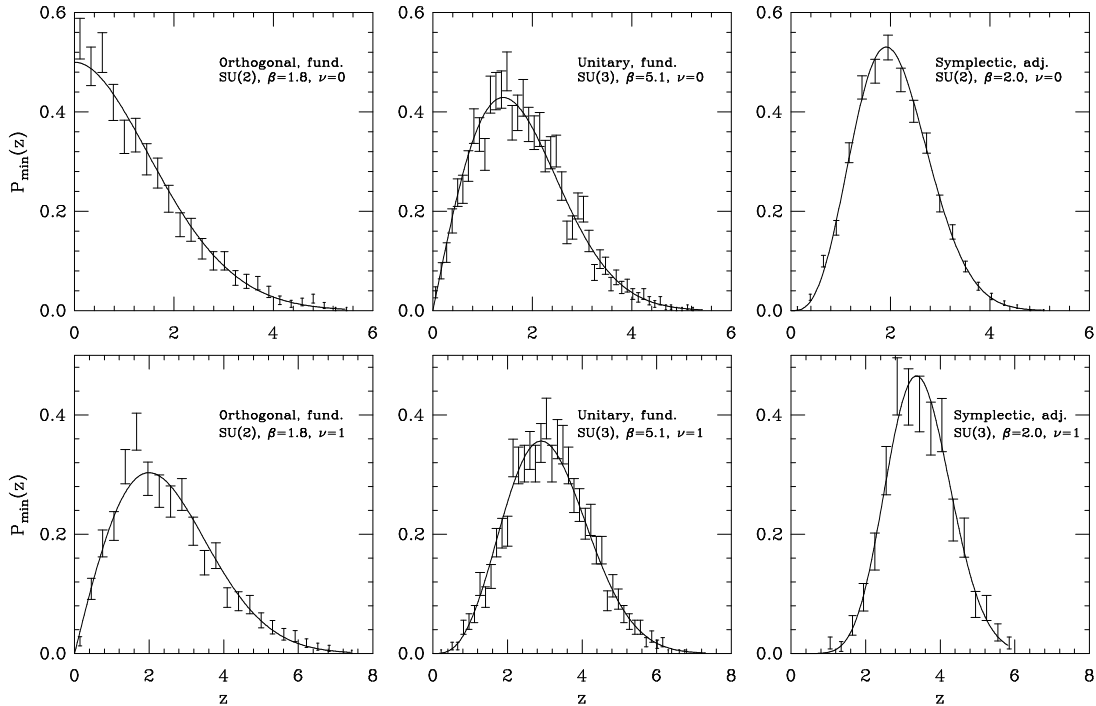}}
\end{center}
\caption{
Plots of $P_{\rm min}(z)$ versus $z$
for the various ensembles in the
lowest two topological sectors. The curve in each plot is 
a fit to the prediction
from random matrix theory with the best value for the chiral condensate.
}
\label{fig:pmin}
\end{figure*}

The classification according to the three RMT ensembles is connected
to the chiral properties of the fermions~\cite{RMTreview}. A good
non-perturbative regularization of QCD should therefore retain those
chiral properties. Until recently such a regularization was not
known. The next best thing were staggered fermions,
which at least retained a reduced chiral-like symmetry on the
lattice. Indeed, staggered fermions were used to verify predictions of
chiral RMT, albeit with two important shortcomings: (i) staggered
fermions in the fundamental representation of SU(2) have the symmetry
properties of the symplectic ensemble, not the orthogonal ensemble as
continuum fermions, while adjoint staggered fermions belong to the
orthogonal ensemble, not the symplectic one.  (ii) staggered fermions
do not have exact zero modes at finite lattice spacing, even for
topologically non-trivial gauge field backgrounds, and thus seem to
probe only the $\nu=0$ predictions of chiral RMT.

The development of the overlap formalism for chiral fermions on the
lattice~\cite{over} led to the massless overlap Dirac operator,
a lattice regularization for vector-like gauge theories that retains
the chiral properties of continuum fermions on the lattice~\cite{herbert}.
In particular, the continuum predictions of chiral RMT
should apply. Overlap fermions have exact zero modes in topologically
non-trivial gauge field backgrounds~\cite{ehn1}, allowing, for the first
time, verification of the RMT predictions in $\nu \ne 0$ sectors.
The nice agreement we shall describe further validates the chiral RMT
predictions and strengthens the case for the usefulness
of the Overlap regularization of massless fermions.

The massless overlap Dirac operator~\cite{herbert} is given by
\begin{equation}
D={1\over 2} \left [ 1 + \gamma_5 \epsilon (H_w(m)) \right ] ~.
\label{eq:over}
\end{equation}
Here, $\gamma_5 H_w(-m)$ is the usual Wilson-Dirac operator
and $\epsilon$ denotes the sign function. The mass $m$ has
to be chosen to be positive and well above the critical mass for Wilson
fermions but below the mass where the doublers become light.
We are interested in the low lying eigenvalues of the hermitian operator
$H=\gamma_5 D$ described in Ref.~\cite{ehn1}.
We will use the Ritz algorithm~\cite{Ritz} applied to $H^2$ to obtain the
lowest few eigenvalues. The numerical algorithm involves the action of $H$ on
a vector and for this purpose one will have to use a representation of
$\epsilon(H_w(m))$. We used the rational approximation discussed
in Ref.~\cite{ehn1}. 

We computed the distribution of the lowest lying eigenvalue of the
overlap Dirac operator in the fundamental representation on pure gauge
SU(2) configurations with $\beta=1.8$ as an example of the chiral orthogonal
ensemble, on pure gauge SU(3) configurations with $\beta=5.1$ as an example
of the chiral unitary ensemble, and in the adjoint representation on
pure gauge SU(2) configurations with $\beta=2.0$ as an example of the
chiral symplectic ensemble. The lattice size was
$4^4$ in all cases.
Chiral RMT predicts that these distributions are universal
when they are classified according to the three ensembles and according to the
number of exact zero modes $\nu$ within each ensemble and then considered
as functions of the rescaled variable $z=\Sigma V \lambda_{\rm min}$.
Here $V$ is the volume and $\Sigma$ is the infinite volume value of the
chiral condensate $\langle \bar \psi \psi \rangle$ 
determined up to an overall wave function normalization, which is dependent 
in part on the Wilson--Dirac mass $m$.
RMT gives the distribution of the rescaled lowest eigenvalue.
A collection of the necessary formulae for the distribution of the
lowest eigenvalue, $P_{\rm min}(z)$, can be found in~\cite{EHNprl}.

We compare the RMT predictions with our data in Fig.~\ref{fig:pmin}.
If $\Sigma$ is known, the RMT predictions for $P_{\rm min}(z)$ are
parameter free. On the rather small systems that we considered here,
we did not obtain direct estimates of $\Sigma$. Instead, we made
one-parameter fits of the measured distributions, obtained from
histograms with jackknife errors, to the RMT predictions,
with $\Sigma$ the free parameter. Our results and some
additional information are given in
Table~\ref{tab:Sigma}. We note the consistency of the values for $\Sigma$
obtained in the $\nu=0$ and $\nu=1$ sectors of each ensemble.
Alternatively, we could have used the value of $\Sigma$ obtained in
the $\nu=0$ sector, to obtain a parameter free prediction for the
distribution of the rescaled lowest eigenvalue in the $\nu=1$
sector. Obviously, the predictions would have come out very well.

With the fermions in the fundamental representation,
we found 81 (for SU(2)), and 147 (for SU(3)) configurations with
two zero modes and 1 and 3 with three zero modes. For the orthogonal
ensemble, we are not aware of a prediction for $P_{\rm min}(z)$ in the
$\nu=2$ sector, while for the unitary ensemble our data, albeit with
very limited statistics, agrees reasonably well with the parameter free
prediction with $\Sigma$ from Table~\ref{tab:Sigma}.

For fermions in the adjoint representation, we keep only one of each
pair of degenerate eigenvalues so $\nu=1$ is the sector with
two exact zero modes. Such configurations cannot
be assigned an integer topological charge since integer charges
give rise to zero modes in multiples of four~\cite{adjoint}, and we
note there are a significant number of configurations with two
zero modes as seen in Table~\ref{tab:Sigma}. The good agreement
with the RMT prediction found in this case lends further support to
the existence of configurations with fractional topological
charge~\cite{adjoint}.

We have tested the predictions of chiral random
matrix theory using the overlap Dirac operator on pure gauge
field ensembles. We find the distribution of the lowest eigenvalue in the
different topological sectors fits well with the predictions of
chiral RMT, with compatible values for the chiral condensate from the
different topological sectors. 

\begin{table}
\caption{The chiral condensate, $\Sigma$, from fits of the distribution
of the lowest eigenvalue to the RMT predictions. The third column gives
the Wilson-Dirac mass parameter used, the fourth the number of
configurations, $N_\nu$, in each topological sector.}
\label{tab:Sigma}
\begin{tabular}{|l|c|c|c|r|c|c|}
\hline
 Repr. & $\beta$ & $m$ & $\nu$ & $N_\nu$ & $\Sigma$ \\
\hline
 SU(2) fund. & 1.8 & 2.3 & 0 & 1293 & 0.2181(51) \\
 SU(2) fund. & 1.8 & 2.3 & 1 & 1125 & 0.2155(37) \\
\hline
 SU(3) fund. & 5.1 & 2.0 & 0 & 2714 & 0.1655(16) \\
 SU(3) fund. & 5.1 & 2.0 & 1 & 2136 & 0.1660(12) \\
\hline
 SU(2) adj.  & 2.0 & 2.3 & 0 & 1251 & 0.2900(30) \\
 SU(2) adj.  & 2.0 & 2.3 & 1 &  254 & 0.2931(45) \\
\hline
\end{tabular}
\end{table}

This research was supported by DOE contracts 
DE-FG05-85ER250000 and DE-FG05-96ER40979.

\end{document}